# Inclusive Photoproduction of Polarized $^3P_1$ Quarkonium

J.P. Ma

Recearch Center for High Energy Physics
School of Physics
University of Melbourne
Parkville, Victoria 3052
Australia

**Abstract**:

We analyse inclusive photoproduction of polarized $^3P_1$ quarkonium in the framework of QCD. To separate nonperturbative and perturbative parts in the density matrix of the produced quarkonium we use a method , which is equivalent to the diagramatic expansion widely used in analysing deeply inelastic scatterings. A systematic expansion in a small velocity $v$, with which a heavy quark moves inside the quarkonium in its rest frame, is performed for the nonperturbative parts, and they are expressed as matrix elements in nonrelativistic QCD. At the leading order of $v$ there are four matrix elements representing nonperturbative physics. The perturbative parts are calculated at the leading order of coupling constants. Some numerical results, especially, numerical results for HERA are given.

## 1. Introduction

A quarkonium mainly consists of a heavy quark $Q$ and its antiquark $\bar{Q}$. The quark $Q$ or the antiquark $\bar{Q}$ carries roughly the half of the momentum of the quarkonium. In a quarkonium rest frame the quark $Q$ or the antiquark $\bar{Q}$ moves with a small velocity $v$. Production of a quarkonium in high energy processes can be considered as a two-step process, at first, a $Q\bar{Q}$ pair is produced, then a transition of the $Q\bar{Q}$ pair into the quarkonium takes place. The production of the $Q\bar{Q}$ pair can be handled perturbatively in QCD because the mass $M$ of $Q$ is large, while the transition is a nonperturbative process. Using an expansion in the small velocity $v$ the nonperturbative process can be described by some constants defined in QCD as matrix elements.

Earlier calculations of quarkonium production have been made only within models. One of them is the color-singlet model. In this model the basic assumptions are that only a color-singlet $Q\bar{Q}$ pair can be transimitted into a quarkonium and the transition is represented by a wavefunction of the color-singlet $Q\bar{Q}$ pair. With these assumptions an important fact in relativistic quantum field theory is neglected, namely a quarkonium as a bound state consists not only of a color-singlet $Q\bar{Q}$ pair, but also of many other states as components. Neglecting this fact leads to that there are infrared singularities in the *perturbative* results related to a $P$-wave quarkonium. These singularities are well known as the divergence in the limit of the zero-binding energy[1]. It is pointed out that a color-octet $Q\bar{Q}$ pair combined with a soft gluon is a component of a $P$-wave quarkonium and this component is as important as a color-sinlet $Q\bar{Q}$ pair[2]. A factorized formula for production rate of an unpolarized $P$-wave quarkonium is given[3], where the nonperturbative transition of a color-singlet and a color-octet $Q\bar{Q}$ pair into the $P$-wave quarkonium are described at the leading order of $v$ respectively by two matrix elements defined in NRQCD. With the factorized formula it is already shown that there is no infrared singularity in the case of gluon fragmentation[4,5].

In this work we consider inclusive photoproduction of single $^3P_1$ quarkonium, where the quarkonium is polarized. The polarization is described by a density matrix. We calculate the density matrix at leading order of the coupling constants and of $v$. To separate nonperturbative effects from perturbative effects we use a method, which is equivalent to the diagramatic expansion widely used in deeply inelastic processes[6–10]. The expansion provides a powerful tool to analyse and to account higher twist contributions in a deeply inelastic process. Although we work here at the leading order of $v$, one can extend with the method our results to higher orders. Some of our results are universal, in the sense that they do not depend on the processes and the quarkonium considered here. Our results for the density matrix show that instead of two for unpolarized case there are four nonperturbative constants for polarized case, three of them are for the color-octet $Q\bar{Q}$ pair. For unpolarized case we rediscover the factorized formula given in [3].

We will denote a $^3P_1$ quarkonium as $\chi_1$. For $c$-quark or $b$-quark it is respectively $\chi_{c1}$ or $\chi_{b1}$. The main decay of $\chi_1$ is radiative decay, where the decay products are one photon and a $S$-wave quarkonium. This decay mode can be used to analyse the polarization information from the production. We will discuss how to construct spin observables for this purpose. Numerical results are given for photoproduction and for HERA, where the photon in the initial state is obtained through equivalent photon approximation from the initial electron at HERA.

Our work is organized as following: In Sect.2 we introduce the density matrix and separate in the density matrix the nonperturbative part from the perturbative part. In Sect.3 we systematically make the small velocity expansion for the nonperturbative part with NRQCD and give the leading order results. Following the way explored here it is easy to extend the results to higher orders. Sect.4 contains perturbative results for density matrices of various partonic processes at leading order of coupling constants. We discuss first an extension of Furry thoerem. With this the contribution from a color-singlet $Q\bar{Q}$ can be shown to be zero at the considered order. Some numerical results are given in Sect.5. We have not performed an extensive study of different observables, only some simple observables are studied here. In Sect.6 is the summary of our work. For completeness we give in an appendix the density matrix for the radiative decay of $\chi_1$.

## 2. The density Matrix and the Factorization

We consider the following inclusive process:

$$\gamma(q) + A(p) \to H(P_H) + X \tag{2.1}$$

where $X$ denotes the unobserved states, $H$ is a quarkonium with spin 1 and $A$ is a hadron in the initial state. The momenta are indicated in the brackets and the moving direction of the photon is opposite to the moving direction of $A$. We will assume that the factorization holds for the initial state and work only at the leading twist order, i.e., we will use the parton model for the initial hadron $A$. Therefore, we only need to study the following partonic process:

$$\gamma(q) + a(q_1) \to H(P_H) + X \tag{2.2}$$

where $a$ stands for a parton with $\mathbf{q_1} = x\mathbf{p}$. In (2.2) the polarization information of the quarkonium is contained in a density matrix $R$, which can be defined as:

$$R(\lambda, \lambda', q, q_1, P_H) = \sum{}' \sum_X (2\pi)^4 \delta^4(q + q_1 - P_H - P_X) \\ \cdot < H(P_H, \lambda), X | \mathcal{T} | \gamma(q), a(q_1) > \\ \cdot < H(P_H, \lambda'), X | \mathcal{T} | \gamma(q), a(q_1) >^* . \quad (2.3)$$

Here $\mathcal{T}$ is the transition operator, $\lambda$ and $\lambda'$ denote the helicities of $H$. $\sum{}'$ means the average of the initial state. In (2.3) the density matrix is given in the helicity basis. Since helicity is not conserved, the density matrix is not diagonal in the helicity basis. Another useful basis is the Cartesian basis defined in the quarkonium rest frame. Throughout our work we define the quarkonium rest frame which is related to the quarkonium moving frame in (2.1) only through a Lorentz boost without rotations. Denoting the polarization vector as $\epsilon(\lambda)$ in the rest frame the density matrix $R_{ij}$ in the Cartesian basis is related to $R(\lambda, \lambda')$ as

$$R(\lambda, \lambda', q, q_1, p_H) = \epsilon_i^*(\lambda) R_{ij}(q, q_1, p_H) \epsilon_j(\lambda'). \quad (2.4)$$

Taking the Lagrangian of QED and QCD, and using Wick Thoerem one can evaluate the amplitude. At certain leading order of the coupling constants the amplitude takes the form:

$$< H(P_H, \lambda), X | \mathcal{T} | \gamma(q), a(q) > = \int \frac{dp_1^4}{(2\pi)^4} A_{ij}(p_1, k_1) \\ \cdot \int dx^4 e^{-ip_1 x} < H(P_H, \lambda), X_N | \bar{Q}_i(x) Q_j(0) | 0 > \quad (2.5) \\ + \cdots$$

where $Q(x)$ is the Dirac field for the heavy quark $Q$, the indices $i, j$ contain the Dirac and color indices. The operators in the matrix element are norm ordered. The matrix element is a nonperturbative object, while the quantity $A_{ij}(p_1, k_1)$ is calculated by perturbative theory. We divided in (2.5) the unobserved state $X$ as $X = X_P + X_N$. The state $X_N$ is generated nonperturbatively and $X_P$ is generated perturbatively with a total momentum as $k_1$. The quantity $A_{ij}(p_1, k_1)$ can be obtained simply by using usual Feynman rule, and it can be interpreted as the amplitude for $\gamma(q) + a(q_1) \to Q^*(p_1) + \bar{Q}^*(q+q_1-p_1-k_1) + X_P(k_1)$. The real amplitude for on-shell $Q$ and $\bar{Q}$ is $\bar{u} A v$ where the on-shell condition is taken. The $\cdots$ in (2.5) denotes the other possible terms by using Wick theorem. For example, there are terms where the matrix element contains only one operator. These terms can be regarded in some circumstances as contributions from parton fragmentations. The fragmentations can be determined by perturbative theory[ 4,5 and references therein], they will eventually take after applying the procedure here for (2.5) the form as in (2.5) with the matrix element containing the product $\bar{Q}_i(x)Q_j(0)$. We will only take the contribution indicated explicitly in (2.5) for the density matrix in (2.3). Substituting (2.5) into (2.3) and using translational transformation to eliminate the summation over $X_N$ we obtain:

$$R(\lambda, \lambda', q, q_1, P_H) = {\sum}' \sum_{X_P} \int \frac{d^4p_1}{(2\pi)^4} \frac{d^4p'_1}{(2\pi)^4}$$
$$A_{ij}(p_1, k_1)\left(\gamma_0 A^\dagger(p_1, k_1)\gamma_0\right)_{kl}$$
$$\cdot \int d^4x d^4x' d^4y e^{-iy(q+q_1-k_1)-ip_1 x + ip'_1 x'} \quad (2.6)$$
$$< 0|\bar{Q}_k(-\frac{1}{2}y)Q_l(x' - \frac{1}{2}y)a_H^\dagger(\lambda')a_H(\lambda)\bar{Q}_i(x + \frac{1}{2}y)Q_j(\frac{1}{2}y)|0 >$$

where $a_H^\dagger$ is the create operator for $H$. With the form in (2.6) the contributions to the density matrix can be presented by the Feynman diagram in Fig.1. In Fig.1 the black box is corresponding to the nonperturbative part:

$$\Gamma_{kl,ij}(p_1, k_1) = \int d^4x d^4x' d^4y e^{-iy(q+q_1-k_1)-ip_1 x + ip'_1 x'}$$
$$< 0|\bar{Q}_k(-\frac{1}{2}y)Q_l(x' - \frac{1}{2}y)a_H^\dagger(\lambda')a_H(\lambda)\bar{Q}_i(x + \frac{1}{2}y)Q_j(\frac{1}{2}y)|0 >, \quad (2.7)$$

while the blank part is a perturbative part and can be obtained by usual Feynman rules except the two pair of quark-lines connecting to the black box. These lines only indicate momentum flow and certain contractions of Dirac and color indices. The diagram in Fig.1 is similar to those appearing in using diagramatic expansion for analysing deeply inelastic processes.

The expression in (2.6) can be simplified by decomposing the color and Dirac indices. The decomposition of the color indices can be easily done with $SU(3)$ color-symmetry. We

obtain the following:

$$R(\lambda, \lambda', q, q_1, P_H) = R^{(1)}(\lambda, \lambda', q, q_1, P_H) + R^{(8)}(\lambda, \lambda', q, q_1, P_H),$$

$$R^{(1)}(\lambda, \lambda', q, q_1, P_H) = \frac{1}{9} {\sum}' \sum_{X_P} \int \frac{d^4 p_1}{(2\pi)^4} \frac{d^4 p_1'}{(2\pi)^4} \text{Tr}_{\text{color}} A_{ij}(p_1, k_1)$$

$$\cdot \text{Tr}_{\text{color}}(\gamma_0 A^\dagger(p_1, k_1)\gamma_0)_{kl} \cdot \Gamma^{(1)}_{kl,ij}(p_1, k_1),$$

$$R^{(8)}(\lambda, \lambda', q, q_1, P_H) = \frac{1}{2} {\sum}' \sum_{X_P} \int \frac{d^4 p_1}{(2\pi)^4} \frac{d^4 p_1'}{(2\pi)^4} \text{Tr}_{\text{color}} T^a A_{ij}(p_1, k_1)$$

$$\cdot \text{Tr}_{\text{color}} T^a (\gamma_0 A^\dagger(p_1, k_1)\gamma_0)_{kl} \cdot \Gamma^{(8)}_{kl,ij}(p_1, k_1),$$

$$\Gamma^{(1)}_{kl,ij}(p_1, k_1) = \int d^4 x\, d^4 x'\, d^4 y\, e^{-iy(q+q_1-k_1)-ip_1 x + ip_1' x'}$$

$$< 0 | \bar{Q}_k(-\frac{1}{2}y) Q_l(x' - \frac{1}{2}y) a_H^\dagger(\lambda') a_H(\lambda) \bar{Q}_i(x + \frac{1}{2}y) Q_j(\frac{1}{2}y) | 0 >,$$

$$\Gamma^{(8)}_{kl,ij}(p_1, k_1) = \int d^4 x\, d^4 x'\, d^4 y\, e^{-iy(q+q_1-k_1)-ip_1 x + ip_1' x'}$$

$$< 0 | \bar{Q}_k(-\frac{1}{2}y) T^b Q_l(x' - \frac{1}{2}y) a_H^\dagger(\lambda') a_H(\lambda) \bar{Q}_i(x + \frac{1}{2}y) T^b Q_j(\frac{1}{2}y) | 0 >.$$

(2.8)

Here $T^a (a = 1, \cdots 8)$ are the $SU(3)$ color-matrices. The indices $i, j, k$ and $l$ now are only Dirac indices. In (2.8) there is a contribution $R^{(8)}(\lambda, \lambda', q, q_1, P_H)$ from a color-octet $Q\bar{Q}$ pair and also a contribution $R^{(1)}(\lambda, \lambda', q, q_1, P_H)$ from a color-singlet $Q\bar{Q}$. We will call them color-octet and color-singlet components respectively. The quantity $\Gamma^{(1)}_{kl,ij}(p_1, k_1)$ and $\Gamma^{(8)}_{kl,ij}(p_1, k_1)$ describe the nonperturbative transition from a color-singlet $Q\bar{Q}$ and from a color-octet $Q\bar{Q}$ pair into the quarkonium. In the color-singlet model it is assumed that $\Gamma^{(8)}_{kl,ij}(p_1, k_1) = 0$.

We will make the small velocity expansion for $\Gamma^{(1)}$ and $\Gamma^{(8)}$ before decomposing the Dirac indices. After the expansion the decomposition can be easily performed. To make the expansion we note the fact discussed in the beginning of the introduction, this fact leads to that the dominant space-time dependence of the matrix elements in (2.8) is given by

$$< 0 | \bar{Q}(-\frac{1}{2}y) Q(x' - \frac{1}{2}y) a_H^\dagger(\lambda') a_H(\lambda) \bar{Q}(x + \frac{1}{2}y) Q(\frac{1}{2}y) | 0 >$$

$$\sim \exp\{ -i\frac{1}{2} P_H(x' - \frac{1}{2}y - \frac{1}{2}y) + i\frac{1}{2} P_H(x + \frac{1}{2}y + \frac{1}{2}y)\}.$$

(2.9)

In the quarkonium rest frame, the matrix element has only a weak dependence on **x**, **x**′, and **y**, this dependence is controlled by the small scale $Mv$, while the dominant time-dependence is given by (2.9) and the correction to this is controlled by the scale $Mv^2$.

The difference between the space- and time-dependences naturely leads to use NRQCD to make the expansion in $v$. The another advantage of using NRQCD is due the recent progress on lattice study, where nonperturbative properties of a quarkonium can be well studied with lattice NRQCD[11].

### 3. The Small Velocity Expansion

In this section we will work in the quarkonium rest frame. To avoid confusions due to too many notations and bookings we still use the same booking for momenta as in Sect.2. Finally we will express our results in a covariant form. In the expansion we will have space-time derivative $\partial_\mu$, we will automatically substitute it with the covariant derivative $D_\mu$, since this is the plausible way to maintain the gauge invariance. Before going to the expansion we discuss briefly how the expansion of the original QCD Lagrangian in the power of $v$ to obtain NRQCD.

We write the Dirac field $Q(x)$ as

$$Q(x) = e^{-iMt}Q_+(x) + e^{iMt}Q_-(x), \tag{3.1}$$

as discussed in the last section, the four component field $Q_+(x)$ and $Q_-(x)$ have a weak dependence on $x$. With the equation of the motion these fields take form:

$$\begin{aligned} Q_+(x) &= P_+ h_+(x) - \frac{i\gamma_i D_i}{2M} P_+ h_+(x) + O(v^2) \\ Q_-(x) &= P_- h_-(x) - \frac{i\gamma_i D_i}{2M} P_- h_-(x) + O(v^2), \end{aligned} \tag{3.2}$$

here $P_\pm = (1 \pm \gamma_0)/2$, $h_+(x)$ and $h_-(x)$ are certain four component fields. Since there are always the combination $P_- h_-(x)$ and $P_+ h_+(x)$ appearing in (3.2), we use two two-component fields $\psi$ and $\chi$ to denote $P_+ h_+(x)$ and $P_- h_-(x)$. The Lagrangian for the two-component fields upto $v^2$ can be easily obtained, which is the Lagrangian of NRQCD:

$$L_{\rm NRQCD} = \psi^\dagger (iD_0 + \frac{D_i^2}{2M})\psi - \chi^\dagger (iD_0 - \frac{D_i^2}{2M})\chi \tag{3.3}$$

The physical interpretation for the two-component fields is: the field $\psi$ destructs a heavy quark while the field $\chi$ creates a heavy antiquark. $L_{\rm NRQCD}$ has the exact spin-symmetry. However, higher order corrections to $L_{\rm NRQCD}$ will violate this symmetry.

### 3.1 The Color-Octet Part $\Gamma^{(8)}$

At leading order of $v$ we only need to substitute (3.1) and (3.2) into $\Gamma^{(8)}$, where the space-time dependence of $P_-h_-(x)$ and $P_+h_+(x)$ should be neglected:

$$\Gamma^{(8)}_{kl,ij}(p1,k_1) = -\int d^4x d^4x' d^4y \exp\{-iy(q+q1-k_1) - ip_1x + ip'_1x' + iP(2y-x'+x)\}$$
$$\cdot <0|(h^\dagger_-P_-)_k T^a(P_+h_+)_l a^\dagger_H(\lambda')a_H(\lambda)(h^\dagger_+P_+)_i T^a(P_-h_-)_j|0>.$$
(3.4)

Here the fields are at the space-time origin, the momentum $P$ is $(M,0,0,0)$. At leading order of $v$ the difference $M_H - 2M$, which is at order $v^2$, can also be neglected(we will discuss this later), i.e., we can take $P_H = 2P$. Since only local operators are involved in (3.4), the Dirac indices can be now decomposed easily. We obtain:

$$\Gamma^{(8)}_{kl,ij}(p1,k_1) = -(2\pi)^{12}\delta^4(q+q_1-k_1-P_H)\delta^4(p'_1-P)\delta^4(p_1-P)$$
$$\frac{1}{4}\Big\{ \begin{pmatrix} 0 & I \\ 0 & 0 \end{pmatrix}_{lk} \begin{pmatrix} 0 & 0 \\ I & 0 \end{pmatrix}_{ji} i <0|\chi^\dagger T^a\psi a^\dagger_H(\lambda')a_H(\lambda)\psi^\dagger T^a\chi|0>$$
$$+ \begin{pmatrix} 0 & \sigma_{l_1} \\ 0 & 0 \end{pmatrix}_{lk} \begin{pmatrix} 0 & 0 \\ \sigma_{l_2} & 0 \end{pmatrix}_{ji} <0|\chi^\dagger \sigma_{l_1} T^a\psi a^\dagger_H(\lambda')a_H(\lambda)\psi^\dagger T^a \sigma_{l_2}\chi|0>\Big\}.$$
(3.5)

Here $\sigma_i(i=1,2,3)$ are Pauli matrices. The $\delta$ functions will force the quark $Q$ and $\bar{Q}$ in the amplitude $A(p_1,k_1)$ to be on-shell. The physical meaning of the matrix elements is also clear, the one without Pauli matrices represents the transition of a $^1S_0$ $Q\bar{Q}$ pair into the quarkonium, the other one with Pauli matrices represents the transition of a $^3S_1$ $Q\bar{Q}$ pair into the quarkonium. Now we take the case in which the quarkonium is a $^3P_1$ quarkonium. As already pointed the transition of a $^3S_1$ $Q\bar{Q}$ pair has a probability proportional to $v^2$[2], while the transition of a $^1S_0$ $Q\bar{Q}$ can only happen at higher order than $v^2$. This can be easily seen with arguments from perturbative theory. Since the spin configuration of a $^1S_0$ $Q\bar{Q}$ is not the same as the $^3P_1$ quarkonium, a gluon emission to restore the configuration is needed and such emission must be due to the interaction part which violates the spin-symmetry in NRQCD. This part is proportional to $v^3$ at least. Therefore the probability of the transition of a $^1S_0$ $Q\bar{Q}$ pair into a $^3P_1$ quarkonium is at least at the order $v^6$. Hence at leading order of $v$ only a $^3S_1$ $Q\bar{Q}$ pair contributes.

The matrix element $<0|\chi^\dagger\sigma_{l_1}T^a\psi a^\dagger_H(\lambda')a_H(\lambda)\psi^\dagger T^a\sigma_{l_2}\chi|0>$ can be simplified with rotation invariance. We write the creation operator for $\chi_1$ as:

$$a^\dagger_H(\lambda) = -\varepsilon_i(\lambda)a^\dagger_i. \tag{3.6}$$

With the operator $a_i^\dagger$ the matrix element can be written:

$$< 0|\chi^\dagger \sigma^j T^a \psi a_H^\dagger(\lambda') a_H(\lambda) \psi^\dagger T^a \sigma^i \chi|0> = \frac{1}{3}\delta_{ij}\delta_{\lambda\lambda'} a^{(8)}$$
$$+ \left(\epsilon_i^*(\lambda)\epsilon_j(\lambda') - \epsilon_j^*(\lambda)\epsilon_i(\lambda')\right) b^{(8)}$$
$$+ \left(\epsilon_i^*(\lambda)\epsilon_j(\lambda') + \epsilon_j^*(\lambda)\epsilon_i(\lambda') - \frac{2}{3}\delta_{ij}\delta_{\lambda\lambda'}\right) c^{(8)} \quad (3.7)$$

and

$$a^{(8)} = \frac{1}{3} < 0|\chi^\dagger \sigma_i T^a \psi a_l^\dagger a_l \psi^\dagger T^a \sigma_i \chi|0>$$
$$b^{(8)} = \frac{1}{12} < 0|\chi^\dagger \sigma_i T^a \psi (a_i^\dagger a_j - a_j^\dagger a_i) \psi^\dagger T^a \sigma_j \chi|0> \quad (3.8)$$
$$c^{(8)} = \frac{1}{6} < |\chi^\dagger \sigma_i T^a \psi (a_i^\dagger a_j + a_j^\dagger a_i - \frac{2}{3}\delta_{ij} a_l^\dagger a_l) \psi^\dagger T^a \sigma_j \chi|0>$$

The constants, i.e., the matrix elements in (3.8) are real because of the time-reversal invariance of QCD. Now we are in the position to define a covariant amplitude for generating a color-octet $^3S_1$ $Q\bar{Q}$ pair:

$$T_H^{(8)}(\lambda) = \text{Tr} A(P, k_1) T^a \gamma \cdot \epsilon^*(\lambda) \frac{\gamma \cdot P_H + M_H}{2M_H} \quad (3.9)$$

where $M_H = 2M$ and $P_H = 2P$. The trace is taken for Dirac index and color index. With this amplitude the color-octet component is:

$$R^{(8)}(\lambda, \lambda', q, q_1, P_H) = \sum{}' \sum_{X_P} (2\pi)^4 \delta^4(q + q_1 - k_1 - P_H)$$
$$\{\frac{1}{3} a^{(8)} \delta_{\lambda\lambda'} \sum_{\lambda''} |T_H^{(8)}(\lambda'')|^2$$
$$+ b^{(8)} [T_H^{(8)}(\lambda) T_H^{(8)*}(\lambda') - T_H^{(8)}(\lambda') T_H^{(8)*}(\lambda)] \quad (3.10)$$
$$+ c^{(8)} [T_H^{(8)}(\lambda) T_H^{(8)*}(\lambda') + T_H^{(8)}(\lambda') T_H^{(8)*}(\lambda)$$
$$- \frac{2}{3} \delta_{\lambda\lambda'} \sum_{\lambda''} |T_H^{(8)}(\lambda'')|^2]\}$$

This form is a covariant form. The summation over the color index $a$ in (3.10) is understood. There are three nonperturbative constants as NRQCD matrix elements in the color-octet component of the density matrix. No simple relation can be found between them. To go beyond the leading order one should simply take the space-time dependence of quark fields into account, except this one should also include the correction from the binding energy to obtain correct kinematic. At the leading order we have taken $M_H = 2M$. For the quarkonium with energy $E_H$, the binding energy leads to a correction proportional to $M_H^2 v^2/E_H^2$. It is interesting to note that this effect may be included by substituting $M$ with $\frac{1}{2}M_H$ in (3.1), and one obtains then instead of NRQCD an effective theory with $M_H$ as a parameter. However, with such an expansion, $Q$ and $\bar{Q}$ will be not on-shell in the amplitude $A$, it may violate gauge invariances. We do not push the issue further and leave it for future study.

After summation over the polarization only the term with $a^{(8)}$ leads to contributions to the unpolarized cross section. This is corresponding to the factorization formula given in [3], our $a^{(8)}$ defined in (3.8) is related to the matrix element defined in [3] (see notation) as:

$$a^{(8)} = \frac{2M_H}{3} < 0|O_8^{\chi_1}(^3S_1)|0 > . \qquad (3.11)$$

The appearance of $2M_H$ is because the relativistic normalization of the hadron state used here. If parity is conserved and there is no absorptive part in the amplitude $T_H^{(8)}$, the term with $b^{(8)}$ will not contribute to the density matrix.

## 3.2 The Color Singlet Part $\Gamma^{(1)}$

To expand the color-singlet part the vacuum saturation technique will help greatly. In the case of quarkonia this technique is a well controlled approximation[3], with it and (3.1) we obtain:

$$\Gamma^{(1)}_{kl,ij}(p_1,k_1) = \int d^4x d^4x' d^4y \exp\big(-iy(q+q_1-k_1) - ip_1 x + ip'_1 x' + iP(2y + x - x')\big)$$
$$< 0|\bar{Q}_{-k}(-\frac{1}{2}y)Q_{+l}(x' - \frac{1}{2}y)a_H^\dagger(\lambda')|0 >$$
$$< 0|a_H(\lambda)\bar{Q}_{+i}(x + \frac{1}{2}y)Q_{-j}(\frac{1}{2}y)|0 > \cdot \{1 + O(v^4)\}$$
(3.12)

With (3.12) we only need to analyse matrix elments containing two quark fields. Performing a translational transformation we have:

$$< 0|a_H(\lambda)\bar{Q}_{+i}(x + \frac{1}{2}y)Q_{-j}(\frac{1}{2}y)|0 > = < 0|a_H(\lambda)\bar{Q}_{+i}(\frac{1}{2}x)Q_{-j}(-\frac{1}{2}x)|0 > \cdot \{1 + O(v^2)\} \qquad (3.13)$$

The correction in (3.13) is due to the binding energy. Here we can see that at leading order of $v$, our method is equivalent to the method used in the color-singlet model, where one uses wavefunction to project out different states of a color-singlet $Q\bar{Q}$ pair. The analysis of the matrix element in (3.13) is straightforward. For $H$ being a $^3P_1$ quarkonium we expand the matrix element in the r.h.s. of (3.13) by using (3.2) and use the parity-conservation of

QCD. We obtain at the leading order of $v$:

$$<0|a_H(\lambda)\bar{Q}_{+i}(\frac{1}{2}x)Q_{-j}(-\frac{1}{2}x)|0> = -\frac{i}{12}<0|a_{i_1}\psi^\dagger\left(-\frac{i}{2}\overleftrightarrow{D}_{i_2}\sigma_{i_3}\right)\chi|0>\varepsilon_{i_1i_2i_3}$$

$$\left\{\begin{pmatrix} 0 & 0 \\ x_{j_1}\sigma_{j_2}\epsilon^*_{j_3}(\lambda)\varepsilon_{j_1j_2j_3} & 0 \end{pmatrix}_{ji} + \frac{1}{M}\begin{pmatrix} \sigma_{j_1}\epsilon^*_{j_1}(\lambda) & 0 \\ 0 & -\sigma_{j_1}\epsilon^*_{j_1}(\lambda) \end{pmatrix}_{ji}\right\}$$
(3.14)

where $\varepsilon_{i_1i_2i_3}$ is the total antisymmetric symmbol in three dimensions with $\varepsilon_{123}=1$.

To give our results for the color-singlet component we define a nonperturbative constant $a^{(1)}$ and a covariant amplitude for generating a color-singlet $^3P_1$ $Q\bar{Q}$ pair

$$a^{(1)} = \frac{1}{18}|<0|a_{i_1}\psi^\dagger\left(-\frac{i}{2}\overleftrightarrow{D}_{i_2}\sigma_{i_3}\right)\chi|0>\varepsilon_{j_1j_2j_3}|^2,$$

$$T_H^{(1)}(\lambda) = \frac{1}{6\sqrt{2}}\frac{1}{M_H^2}\varepsilon_{\mu\nu\sigma\rho}P_H^\sigma\epsilon^{*\rho}(\lambda)$$

$$\cdot \text{Tr}\left\{\frac{1}{2}A^\mu(P,k_1)\gamma^\nu(\gamma\cdot P_H + M_H) - \frac{1}{M_H}A(P,k_1)\gamma^\mu\gamma\cdot P_H\gamma^\nu\right\},$$
(3.15)

$$A^\mu(P,k_1) = \frac{\partial A(p_1,k_1)}{\partial p_{1\mu}}|_{p_1=P}.$$

With these notataions the color-singlet component of the density matrix is given by

$$R^{(1)}(\lambda,\lambda',q,q_1,P_H) = \sum{}'\sum_{X_P}(2\pi)^4\delta^4(q+q_1-k_1-P_H)$$
$$a^{(1)}\left[T_H^{(1)}(\lambda)(T_H^{(1)}(\lambda'))^*\right]$$
(3.16)

This result is equivalent to the result in the color-singlet model(See [12]), only the non-perturbative effect is expressed with the NRQCD matrix element $a^{(1)}$ instead of a wavefunction. In contrast to the color-octet component we have here only one nonperturbative constant, with the vacuum saturation it is related to the operator given in [3] through:

$$a^{(1)} = \frac{2M_H}{3}<0|O_1^{\chi_1}(^3P_1)|\lambda 0>$$
(3.17)

The constant is proportional to $v^2$ because $D_i$ appears twice in $a^{(1)}$, the color-sinlet component is then at order $v^2$. As we have seen that the color-octet component is also at order $v^2$, hence we need to include both components in a systematical expansion of $v$. With the results in (3.16) and in (3.10), the perturbative effect is in the amplitude $T_H^{(1)}$ and $T_H^{(8)}$, while the nonperturbative effect is included in the four matrix elements defined within NRQCD. These matrix elements should be regarded as defined at a certain renormalization scale, a renormalization group equation for $a^{(8)}$ was derived at one-loop level in [2] and also in [4].

It should be pointed out that similar method was used to analyse quarkonium decay[13], where one only takes the color-singlet $Q\bar{Q}$ and the factorization was performed in the decay amplitude. This is reasonable since in this case the vacuum saturation is a good approximation. However, for the color-octet component the analysis can not be performed only with the amplitude, one needs to consider the density matrix i.e., the probability.

## 4. The Perturbative Part

In this section we will show that $\chi_1$ quarkonium can only be produced in the process in (2.2) at least at the order of $\alpha\alpha_s^2$. Especially, if we neglect heavy quark contents in the initial hadron, only the color-octet components are nonzero at order $\alpha\alpha_s^2$. Before going to calculate the perturbative part, we first discuss an extension of Furry theorem. This extension will enable us to decide which Feynman diagrams to be calculated. Consider a $^3L_J$ quarkonium to be produced via the two diagrams in Fig.2. The only difference between the two diagrams is the direction of the quark line. There are $n$ photons attaching to the quark line, the $i$th photon carries a momentum $k_i$ flowing into the quark loop. These photons may be off-shell or on-shell. Here, only the color-singlet component is involved. The black circle represents the transition of the $Q\bar{Q}$ pair into the quarkonium. For convenience we use a wavefunction $\psi_{LM}$ to describe the transition. Neglecting terms at the order higher than $v$, the contribution from the left diagram in the quarkonium rest frame can be written:

$$A_L = \int \frac{d^4q}{(2\pi)^4} (2\pi)\delta(q^0)\psi_{LM}(\mathbf{q})\Big\{\frac{1}{(p_1-k_1)^2 - M^2} \cdot \frac{1}{(p_1-k_1-k_2)^2 - M^2}$$
$$\cdots \frac{1}{(p_1-k_1-k_2\cdots k_{n-1})^2 - M^2}$$
$$\cdot \text{Tr}\big[\gamma^{\mu_1}(\gamma\cdot(p_1-k_1)+M)\gamma^{\mu_2}(\gamma\cdot(p_1-k_1-k_2)+M)$$
$$\cdots(\gamma\cdot(p_1-k_1-\cdots-k_{n_1})+M)\gamma^{\mu_n}(\gamma\cdot(p_1-P_H)+M)\gamma\cdot\epsilon^*(S_z)(\gamma\cdot p_1+M)\big]\Big\}$$
(4.1)

where $\epsilon^*(S_z)$ is the polarization vector for the total spin of the $Q\bar{Q}$ pair and $p_1 = \frac{1}{2}P_H + q$. To obtain contribution at the leading order of $v$ one only needs to expand in $q$ the terms in the bracket $\{\cdots\}$ in (4.1) and then to perform the integral. One can show that for $L = 0(S$-wave) the leading order contribution is equal to that from the right diagram with a multiplier factor $(-1)^{1+n}$, while for $L = 1(P$-wave) the factor is $(-1)^{2+n}$. We can conclude that for $L = 0(1)$ we do not need to consider the diagrams attached with even(odd) number photons and for the diagrams with odd(even) number photons we only need to calculate the half of them.

Now we consider possible partonic processes in (2.2) which contribute to the color-singlet component of the corresponding density matrix. At leading order of coupling constants, i.e., at $eg_s^2$ in the amplitude, there is one partonic process $\gamma + G \to \chi_1 + G$ via the type of the diagrams in Fig.3, where a $Q\bar{Q}$ color singlet may be produced. In the diagrams the two gluons must have same color, the diagrams are hence QED-like for the color-singlet component. From the above extension of Furry thoerem, we know that the sum of the contributions from the diagrams is zero at leading order of $v$. Therefore the color-sinlet component is zero at this order. In the following we will only consider color-octet component.

The partonic processes below will give at order $\alpha \alpha_s^2$ nonzero contribution to the corresponding color-octet components:

$$\gamma(q) + q(q_1) \to \chi_1(P_H) + q(k_1) \tag{4.2}$$

$$\gamma(q) + \bar{q}(q_1) \to \chi_1(P_H) + \bar{q}(k_1) \tag{4.3}$$

$$\gamma(q) + G(q_1) \to \chi_1(P_H) + G(k_1) \tag{4.4}$$

For these processes we define the usual invariant quantities:

$$\hat{s} = (q + q_1)^2, \quad \hat{t} = (q_1 - P_H)^2. \tag{4.5}$$

The process (4.2) is described by the Feynman diagrams in Fig.4. There are 6 diagrams for the process (4.4), one of them is given in Fig.3, the other five can be obtained through permutation. At first look the diagrams containing the three gluon vertex may lead to generating a color-octet $Q\bar{Q}$ pair, but with the extension of Furry theorem, the sum of these diagrams are zero at leading order of $v$. Further, for the 6 diagrams we need only to calculate half of them and the corresponding amplitude is proportional to $d_{abc}$. We express our results in the Cartesian basis and define three tensors:

$$
\begin{aligned}
C_{1ij} &= q'_i q'_j - \frac{1}{3}\delta_{ij}|\mathbf{q'}|^2 \\
C_{2ij} &= q'_{1i} q'_{1j} - \frac{1}{3}\delta_{ij}|\mathbf{q'_1}|^2 \\
C_{3ij} &= q'_i q'_{1j} + q'_j q'_{1i} - \frac{2}{3}\delta_{ij}\mathbf{q'_1} \cdot \mathbf{q}, \\
\mathbf{q'} &= \mathbf{q} + \mathbf{P_H}\left\{ \left(\frac{P_H^0}{M_H} - 1\right)\frac{\mathbf{P_H} \cdot \mathbf{q}}{|\mathbf{P_H}|^2} - \frac{q^0}{M_H} \right\} \\
\mathbf{q'_1} &= \mathbf{q_1} + \mathbf{P_H}\left\{ \left(\frac{P_H^0}{M_H} - 1\right)\frac{\mathbf{P_H} \cdot \mathbf{q_1}}{|\mathbf{P_H}|^2} - \frac{q_1^0}{M_H} \right\}.
\end{aligned}
\tag{4.6}
$$

Note $\mathbf{q}'$ and $\mathbf{q}'_1$ are momenta in the quarkonium rest frame corresponding to the momenta $\mathbf{q}$ and $\mathbf{q}_1$. With these notations the density matrix for the process (4.2) reads:

$$R_{ij}^q = R_{ij}^{(8),q} = (2\pi)\delta(q_1^0 + q^0 - P_H^0 - k_1^0) \cdot \frac{1}{2k_0} \cdot \frac{e^2 Q_q^2 g_s^4}{96} \cdot \frac{1}{M_H^4 \hat{s}\hat{t}}$$
$$\left\{ \frac{1}{3}\delta_{ij} a^{(8)} \left[ -2M_H^4 + 2M_H^2(\hat{s}+\hat{t}) - \hat{s}^2 - \hat{t}^2 \right] + 2M_H^2 c^{(8)} \left[ C_{1ij} + 2C_{2ij} + C_{3ij} \right] \right\},$$
(4.7)

where the nonperturbative constants $a^{(8)}$ and $c^{(8)}$ are defined in (3.8) and the subscriber $q$ indicales that the initial parton is a quark. $Q_q$ is the electric charge carried by a quark $q$ in the unit of the proton charge. Because of the charge conjungate symmetry the density matrix $R^{\bar{q}}$ for the process (4.3) is the same as given in (4.7). The density matrix for the process (4.4) is:

$$R_{ij}^G = R_{ij}^{(8),G} = (2\pi)\delta(q_1^0 + q^0 - P_H^0 - k_1^0) \cdot \frac{1}{2k_0} \cdot e^2 Q_Q^2 g_s^4 \cdot \frac{40}{3}$$
$$\cdot \frac{1}{(\hat{s}-M_H^2)^2} \cdot \frac{1}{(\hat{t}-M_H^2)^2} \cdot \frac{1}{(\hat{u}-M_H^2)^2}$$
$$\cdot \left\{ \frac{1}{3}\delta_{ij} a^{(8)} \left[ \hat{s}^2(\hat{s}-M_H^2)^2 + \hat{t}^2(\hat{t}-M_H^2)^2 + \hat{u}^2(\hat{u}-M_H^2)^2 \right] \right.$$
$$- 2M_H^2 c^{(8)} \left\{ C_{1ij} \left[ (\hat{s}-M_H^2)^2 + (\hat{u}-M_H^2)^2 - 2M_H^2 \hat{t} \right] \right.$$
$$\left. \left. + C_{2ij} \left[ \hat{s}^2 + \hat{t}^2 \right] + C_{3ij} \hat{s}^2 \right\} \right\}$$
(4.8)

where $\hat{u} = M_H^2 - \hat{s} - \hat{t}$. This result is similar to the result for the process $\gamma + G \rightarrow J/\psi + G$, of which the total cross-section have been calculated in [14]. The differences are in the nonperturbative constants and in the overall color-factors. However, this similarity will be spoiled at higher orders.

With the results in (4.7) and (4.8) we complete the perturbative parts in density matrices for the partonic processes. Before ending this section we would like to point a fact about $\chi_1$ production through gluon fusion $G + G \rightarrow \chi_1$, where only a color-octet $Q\bar{Q}$ pair is produced at first. Landau-Yang theorem does not forbide this process, from the theorem one can only conclude that the amplitude can only be antisymmetric in the color-indices of the two gluons. But the amplitude is accidentally zero at the leading order of $v$[15]. At higher orders it may become nonzero.

## 5. Numerical Results

We will use the radiative decay of $\chi_1$ to analyse the polarization. The whole process we consider is a chain-process:

$$\gamma(q) + A(p) \to \chi_1(P_H) + X$$
$$\chi_1 \to {}^3S_1 + \gamma. \tag{5.1}$$

The photon in the decay carries momentum $k$ in the $\chi_1$ rest frame. The density matrix $\rho_{ij}(\mathbf{k})$ for the decay is given in the appendix. We used the parton model for the initial hadron, hence the momentum $q_1$ in the partonic processes(4.2–4.3) is $q_1 = xp$. Any observable $O$ measured in the process (5.1) can be predicted by:

$$<O> = \frac{1}{\mathcal{N}} \frac{1}{4\pi} \int d\Omega_k \rho_{ji}(\hat{\mathbf{k}}) \int \frac{d^3 P_H}{(2\pi)^3 2P_H^0} \int dx \sum_a \frac{1}{2\hat{s}} f_{a/A}(x) R_{ij}^a \cdot O \tag{5.2}$$

Here $f_{a/A}(x)$ is the parton distribution, $\mathcal{N}$ is a normalization factor so that $<1> = 1$. An observable can be constructed with the three momenta $\mathbf{p}$, $\mathbf{P_H}$ and $\mathbf{k}$. One can also use instead of $\mathbf{k}$ the one observed in the laborotry frame. Any experimental cuts can be implemented into (5.2). These cuts are not only due to experiment for clean signal, but also are necessary in order to select the kinematical region where our perturbative results can be applied. For this purpose we use a simple cut. We require that the transversal momentum of $\chi_1$ to be larger than $P_T$. With sufficient large $P_T$ one can be in confident that the employment of parton model is correct. With this cut the contribution from elastic/diffractive production is discarded, only inelastic production which is handled in this work contributes to the process (5.1). With the cut the factor $\mathcal{N}$ is the total cross-section of unpolarized $\chi_1$. We will take $\chi_1$ to be a charmonium $\chi_{c1}$ and the initial hadron $A$ in (5.1) is a proton. For parton distributions we simply take the HO set given in [16] with $\mu = M_{\chi_{c1}}$. We take $M_H = M_{\chi_{c1}} = 3.51$Gev and $\alpha_s(M_{\chi_{c1}}) = 0.25$. The cut is $P_T = 3$GeV. We consider here the total cross-section $\sigma_{\gamma p}(\chi_{c1})$. The cross-section is proportional to the nonperturbative constant $a^{(8)}$ in our approximation. An estimation of this constant in another convention is given in [2] and in [17]. The relation to the parameter $H_8'$ defined in [2,3] is:

$$a^{(8)} = \frac{1}{2} M_{\chi_{c1}}^3 H_8'(1 + O(v^2)), \tag{5.3}$$

The parameter $H_8'$ estimated in [2,17] lies between 1.4 and 3.0 MeV. We take $H_8' = 2.0$MeV. The numerical result for the total cross section in nb as a function of $\sqrt{s_{\gamma p}} = \sqrt{(q+p)^2}$ in GeV is given in Fig.5. At $P_T = 3$GeV the dominant contribution comes from photon-gluon fusion, while the processes with quark or antiquark as the initial partons contribute only with the fraction at order of $10^{-3}$.

Now we turn to HERA. At HEAR the initial hadron is a proton. The initial photon can be regarded as the one emitted by the initial electron:

$$e \to e + \gamma^*. \tag{5.4}$$

With the equivalent photon approximation(See detail in [18]) this photon in the remainder process like in the process (5.1), can be regarded as real and moving in the same direction as the initial electron. This is just in the same spirit as the parton model, but the photon distribution can be calculated perturbatively, and the distribution is different in different cases. We consider the case without tagging the outcoming electron, the distribution is[19]:

$$f_{\gamma/e} = \frac{\alpha}{\pi}\left\{(1+(1-z)^2)\ln\frac{2E(1-z)}{m_e(2-z)} + \frac{1}{2}z^2\ln\frac{2-z}{z} - (1-z)\right\} + O(\frac{m_e^2}{E^2}) \tag{5.5}$$

where $E$ is the energy carried by the initial electron, $z$ is the energy fraction carried by the photon. The total cross section $\sigma$ at HERA as a function of $P_T$ is drawn in Fig.6. The cross section rapidly decreases as $P_T$ increases. At very large $P_T$ the contribution from the process (4.2) and (4.3), i.e., the initial parton is a quark or antiquark is dominant, and this contribution can be factorized further as a real gluon is first produced and the quarkonium is then produced via gluon fragmentation. But at large $P_T$ $\sigma$ is too small for HERA to observe.

To obtain polarization information of the produced $\chi_{c1}$ a simple way may be to construct some integrated spin observables, where the momentum $k$ of the photon is involved. We construct two of them at HERA

$$O_p = (\hat{\mathbf{k}} \cdot \hat{\mathbf{p}})^2 - \frac{1}{3}, \qquad O_H = (\hat{\mathbf{k}} \cdot \hat{\mathbf{P}}_{\mathbf{H}})^2 - \frac{1}{3} \tag{5.6}$$

where $\hat{\mathbf{k}} = \mathbf{k}/|\mathbf{k}|$, $\hat{\mathbf{p}}$ is the moving direction of the proton and $\hat{\mathbf{P}}_{\mathbf{H}}$ is the moving direction of $\chi_{c1}$ at HERA. Special care must be taken for these observable because they are not Lorentz-invariant. However, it should be noted that the formula in (5.2) is Lorentz-invariant in the sense that $<O>$ is invariant provided $O$ itself is invariant. The expectation values of $O_p$ and $O_H$ are proportional to $c^{(8)}/a^{(8)}$ and no information is available for $c^{(8)}$. The numerical results for $<O_H>$ as a function of $P_T$ are given in Fig.7, where we simply take $c^{(8)} = a^{(8)}$. The expectation value approaches to a constant as $P_T > 8\text{GeV}$. At $P_T = 2\text{GeV}$ we explicitly give our numerical results in the following:

$$\sigma = 0.025\text{nb}, \quad <O_H> = -0.012\frac{c^{(8)}}{a^{(8)}}. \tag{5.7}$$

The expectation value of $O_p$ is much smaller than $<O_H>$. Hence we do not present its results.

## 6. Summary

In this work inclusive photoproduction of polarized $^3P_1$ quarkonium is treated in the framework of QCD. The whole information about the production is contained in density matrices. We have separated in these matrices the nonperturbative and perturbative parts. The nonperturbative parts are worked out at leading order of $v$, at this order there are four matrix elements defined in NRQCD. Three of them represent nonperturbative effects of the transition from a color-octet $Q\bar{Q}$ into $^3P_1$ quarkonium, one is for the transition of a color-singlet $Q\bar{Q}$ pair into $^3P_1$ quarkonium and is corresponding to the wavefunction at the origin in the color-singlet model. With the method employed here one can systematically perform the small velocity expansion and find corrections in higher order in $v$. Although we treated the case of the photoproduction, the factorized form shown here is general and the four matrix elements are universal. The perturbative parts are calculated at leading order of coupling constants. At this order they only receive contributions from color-octet component, i.e., a $Q\bar{Q}$ pair in color-octet is first produced, then this pair is transmitted into a $3_1^P$ quarkonium nonperturbatively. At higher order in $\alpha_s$ the color-singlet component will become nonzero.

Some numerical results are given in this work. We calculated the total cross section of photoproduction and also for HERA. The cross section is not very small for observation. Two spin observables are constructed for HERA, the numerical results for one of them is given. If one can measure these observables one can determine one of these four matrix elements, which is unknown now. Our numerical results are based on leading order results of the density matrices and they can be changed if one adds higher order corrections. In general the corrections from next leading order in $\alpha_s$ are numerically important, also, large corrections from the next leading order of $v$ for charmonium are expected. Due to lack of these corrections we have not performed the numerical study extensively. This can be done once the corrections are known.

## Appendix

In this appendix we consider the radiative decay of $\chi_1$ in its rest frame:

$$\chi_1 \to\, ^3S_1 + \gamma \qquad (A.1)$$

where the photon carries momentum $k$. The mass difference between the two quarkonia is at order of $v^2$. We will assume that the polarization of the $^3S_1$ quarkonium and the photon is not observed. The density matrix $\rho_{ij}$ in this case can be defined in a similar way as (2.3). The decay amplitude can be expanded in $v$. Assuming $C$- and $P$-symmetries to be held, at leading order the amplitude reads[20]:

$$<^3S_1,\gamma|\mathcal{T}|\chi_1> = f\varepsilon_{\mu\nu\alpha\beta}k^\mu \epsilon_\gamma^{*\nu}\epsilon_{^3S_1}^{*\alpha}\epsilon_{\chi_1}^\beta. \qquad (A.2)$$

Here $f$ is a form-factor, other notations are self-explained. With this amplitude we can obtained the density matrix $\rho_{ij}$ at the leading order of $v$:

$$\rho_{ij}(\hat{\mathbf{k}}) = \delta_{ij} + \frac{3}{4}\left(\hat{k}_i\hat{k}_j - \frac{1}{3}\delta_{ij}\right) \qquad (A.3)$$

where $\hat{\mathbf{k}} = \mathbf{k}/|\mathbf{k}|$. The density matrix is normalized:

$$\frac{1}{4\pi}\int d\Omega_k \rho_{ij} = \delta_{ij} \qquad (A.4)$$

where $\Omega_k$ is the solid angle of $\mathbf{k}$.

**Figure Captions**

Fig.1: The Feynman diagram representing Eq.(2.6). The time-direction is from left to right. The broken line is the Cutkosky cut. The black box represents the nonperturbative part while the blank part is the perturbative part.

Fig.2 The Feynman diagrams for quarkonium production

Fig.3 One of the Feynman diagrams for quarkonium production through photon-gluon fusion.

Fig.4 The Feynman diagrams for process $\gamma + q \to Q\bar{Q} + q$.

Fig.5 The total cross section in nb for $\chi_{c1}$ photoproduction as function of $\sqrt{s_{\gamma p}}$ in GeV.

Fig.6 The total cross section in nb for HERA as a function of $P_T$ in GeV.

Fig.7 The expectation value $<O_H>$ for HERA as a function of $P_T$ in GeV.

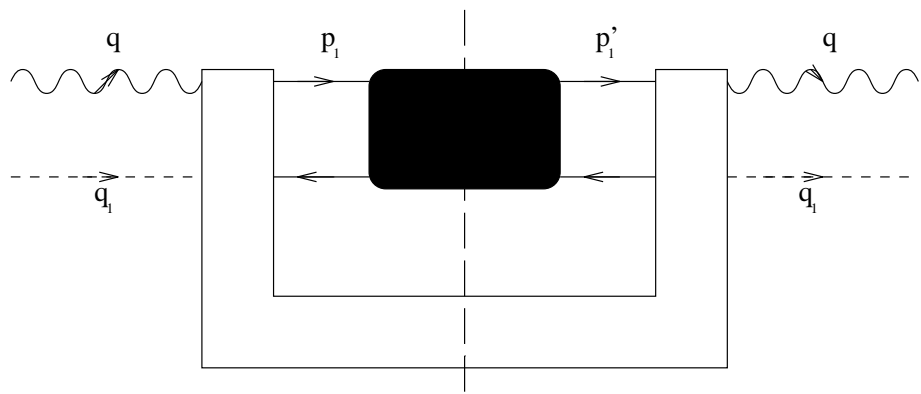

Fig. 1

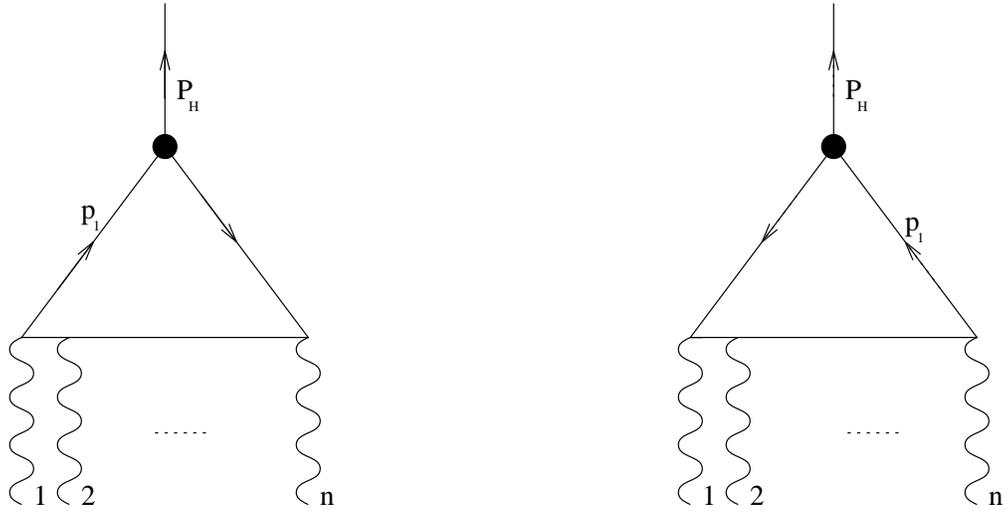

Fig. 2

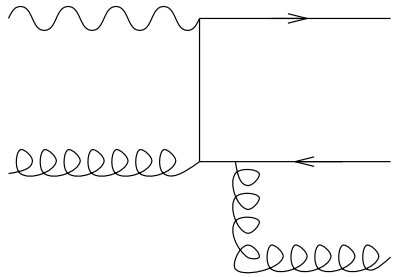

Fig. 3

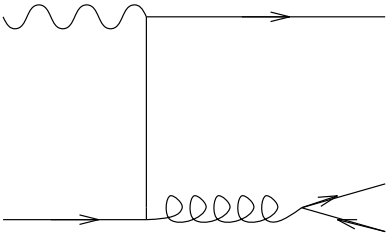 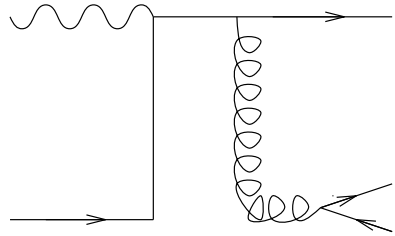

Fig.4

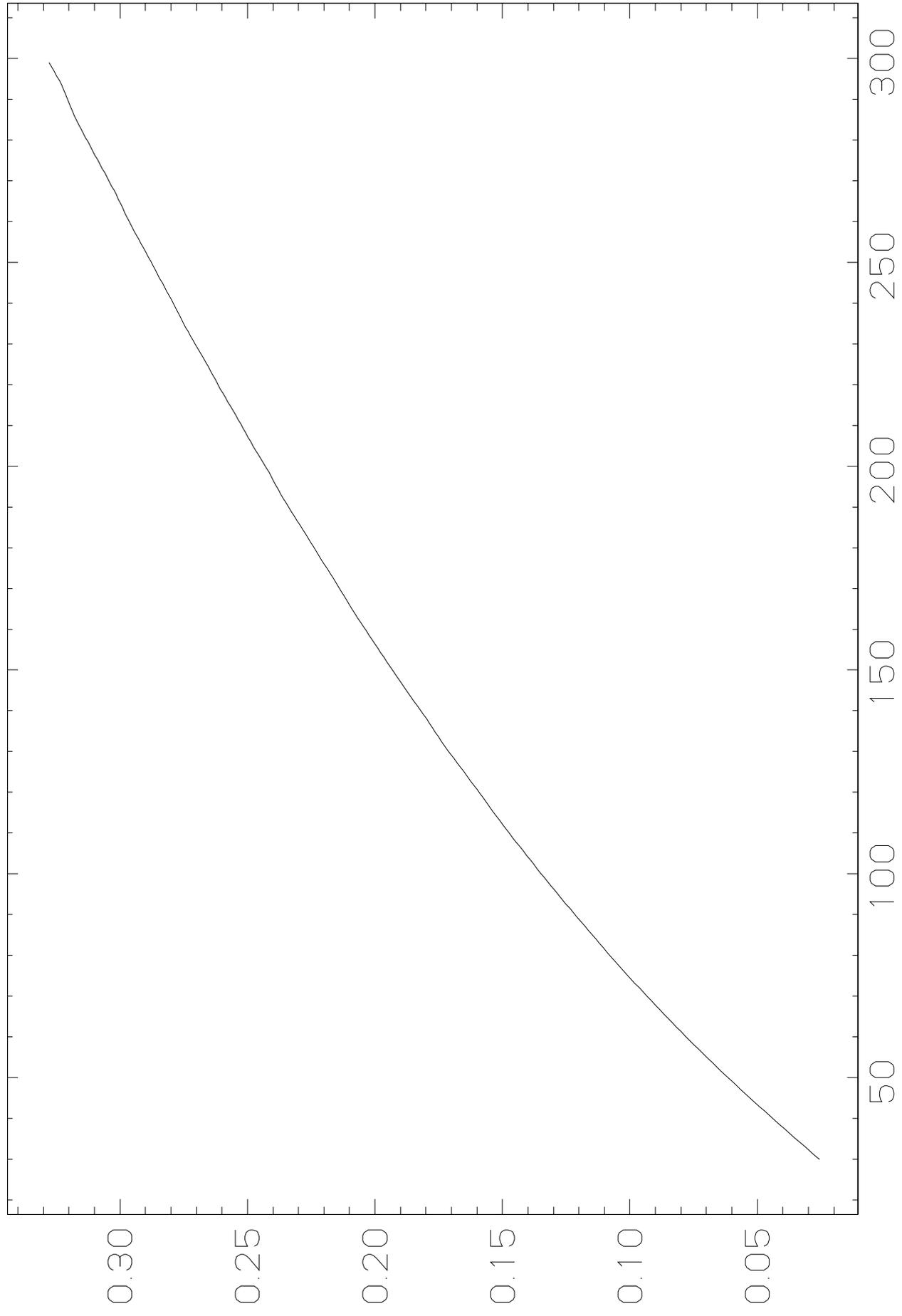

Fig.5

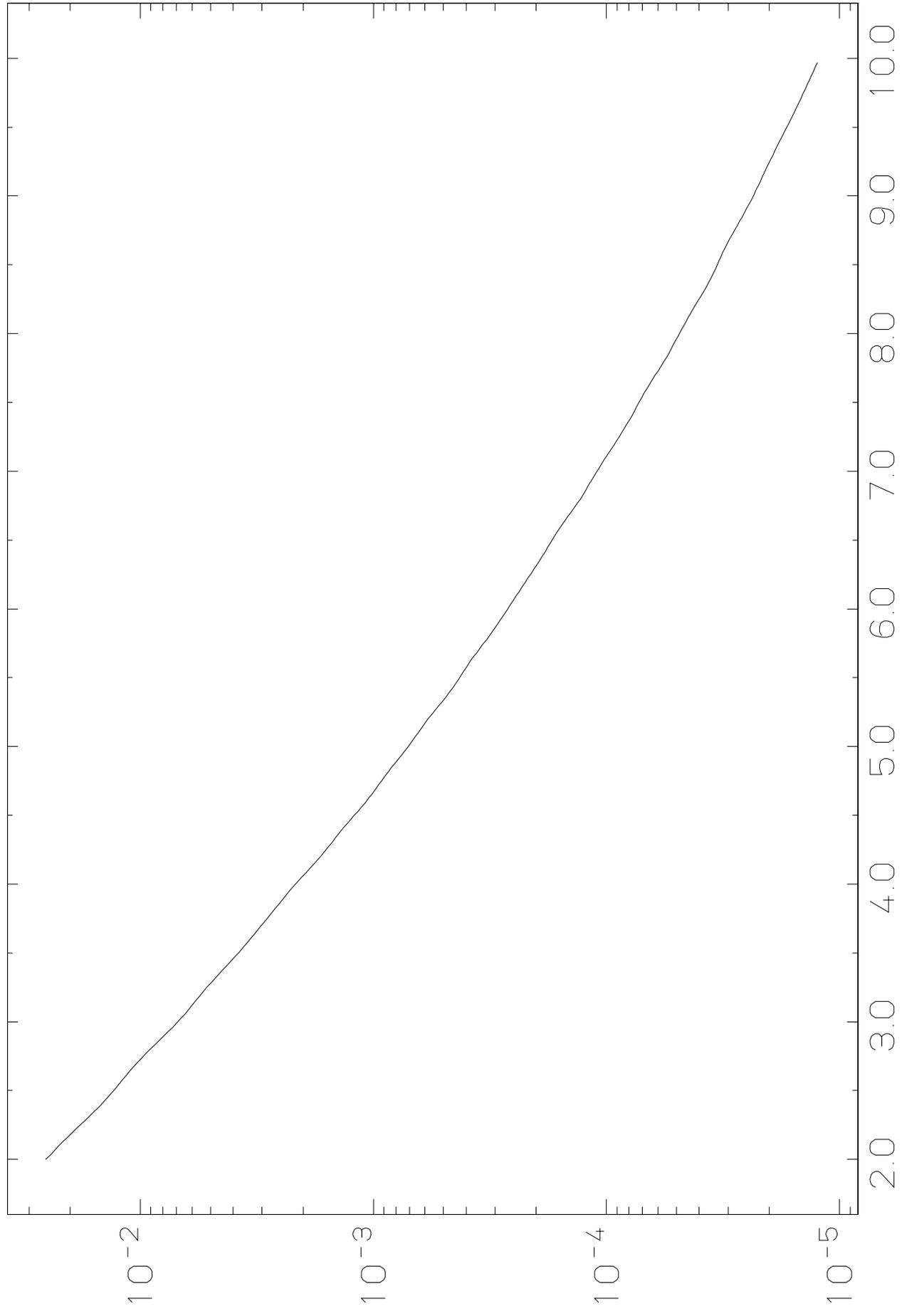

Fig.6

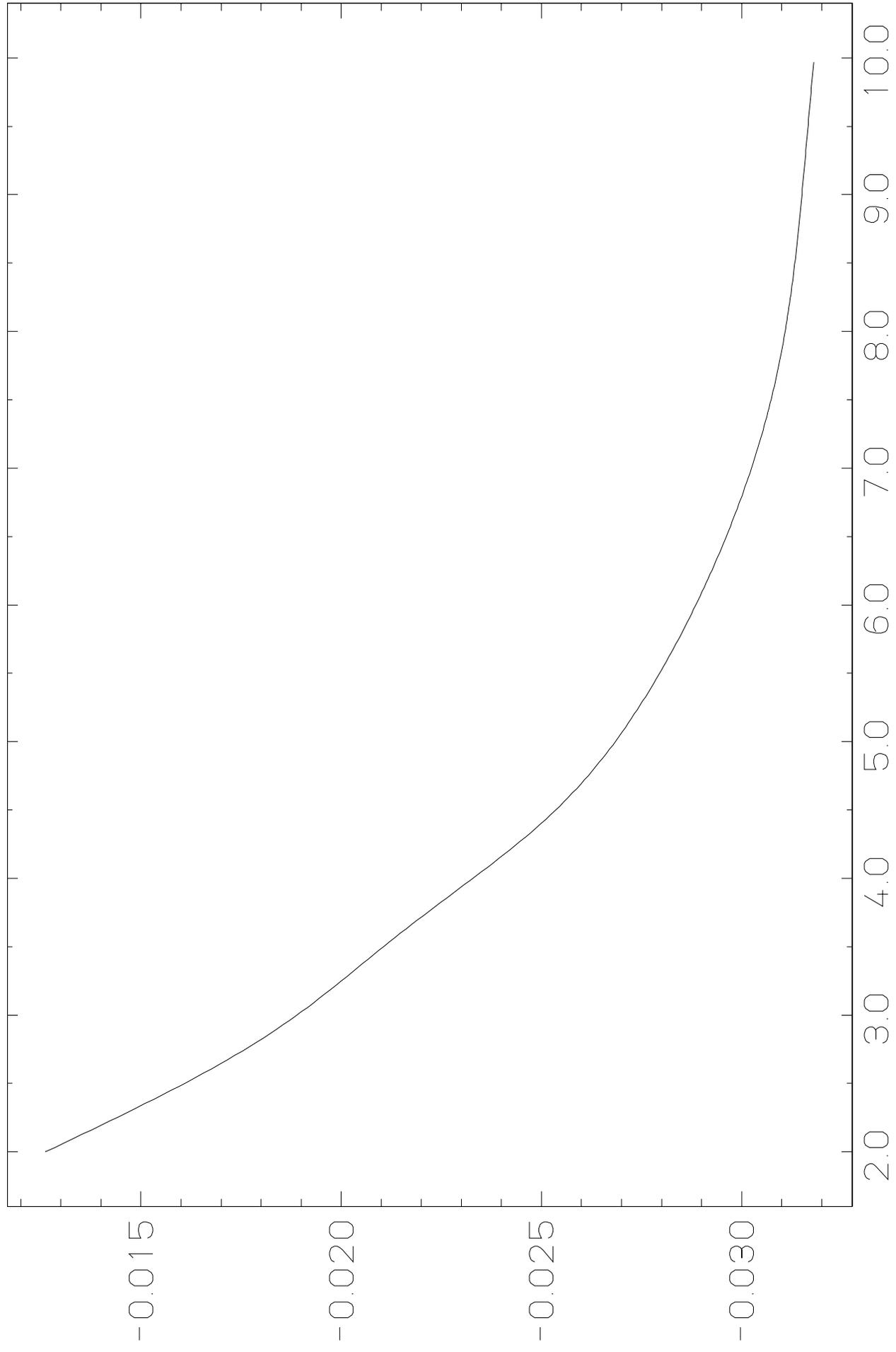

Fig.7